\documentclass[11pt,a4paper]{article}
\pdfoutput=1
\usepackage{jcappub}
\newcommand{\be}{\begin{equation}}
\newcommand{\ee}{\end{equation}}
\newcommand{\bea}{\begin{eqnarray}}
\newcommand{\eea}{\end{eqnarray}}
\newcommand{\bel}[1] {\begin{equation}\label{#1}}
\newcommand{\beal}[1] {\begin{eqnarray}\label{#1}}
\def\({\left(}
\def\){\right)}
\def\[{\left[}
\def\]{\right]}

\def\nn{\nonumber}
\def\be{\begin{equation}}
\def\bea{\begin{eqnarray}}
\def\ee{\end{equation}}
\def\eea{\end{eqnarray}}
\def\nn{\nonumber \\}

\def\bml{\begin{subequations}}
\def\eml{\end{subequations}}

\usepackage[english]{babel}

\begin{document}

\title{ Effects on the CMB from Compactification Before Inflation}

\author[a]{Eleni-Alexandra Kontou}
\affiliation[a]{Physics Program, Bard College, 30 Campus Rd, Annandale-on-Hudson, NY 12504, USA}
\emailAdd{ekontou@bard.edu, elenikontou@cosmos.phy.tufts.edu}
\author[b,c]{Jose J. Blanco-Pillado}
\affiliation[b]{IKERBASQUE, Basque Foundation for Science, 48011, Bilbao, Spain}
\emailAdd{josejuan.blanco@ehu.es}
\affiliation[c]{Department of Theoretical Physics, University of the Basque Country, UPV/EHU, 48080, Bilbao, Spain}
\author[d]{Mark P. Hertzberg}
\emailAdd{ mark.hertzberg@tufts.edu}
\author[d]{Ali Masoumi}
\emailAdd{ali@cosmos.phy.tufts.edu}
\affiliation[d]{Institute of Cosmology, Department of Physics and Astronomy,\\ 
Tufts University, Medford, MA 02155, USA}

\abstract{
Many theories beyond the Standard Model include extra dimensions,  though these have yet to be directly observed. In this work we consider the possibility of a compactification mechanism which both allows extra dimensions and is compatible with current observations. This compactification is predicted to leave a signature on the CMB by altering the amplitude of the low $l$ multipoles,  dependent on the amount of inflation. Recently discovered CMB anomalies at low multipoles may be evidence for this. In our model we assume the spacetime is the product of a four-dimensional spacetime and  flat extra dimensions. Before the compactification, both the four-dimensional spacetime and the extra dimensions can either be expanding or contracting independently. Taking into account physical constraints, we explore the observational consequences and the plausibility of these different models.}

\maketitle

\section{Introduction}

Many of the extensions of the Standard Model involve theories which live 
in a higher dimensional spacetime. However, all current observational
evidence points to a 4-dimensional description of the universe at large scales. A natural way to accommodate these higher dimensional theories
with our observations is to allow for a compactification
mechanism. This is normally achieved by the presence of a higher
dimensional energy-momentum tensor that creates
the conditions for a highly anisotropic evolution of the spacetime. In
particular, our universe today seems to be described well in this
context by a $\mathbb{R}^4\times\mathcal{M}$ cosmological model where the
extra-dimensional manifold $\mathcal{M}$ remains stabilized.
One can then assume that  the compactification mechanism fixes the size of the
internal space to be small enough such that the degrees of freedom 
associated with these extra-dimensions are effectively 
out of reach for the low energy 4-dimensional observers. 

There are however, important open questions in this class of
models. For example, it is hard to argue that our four-dimensional
cosmological spacetime would be the unique vacuum solution in a
higher dimensional setting. So one may 
wonder how our four-dimensional spacetime is dynamically selected.
There are several ideas in the literature that attempt to address these 
questions \cite{Brandenberger:1988aj,Karch:2005yz}, usually invoking extended objects in various dimensions with dynamics that enforces compactification down to three large spatial dimensions.

Another possibility would be if the higher dimensional theory allows
for many vacua with different numbers of compactified dimensions, and our
universe is only one of a large number of possible compactifications.
These kinds of ideas have been recently explored in the context of simple 
higher dimensional field theories where the compactification was obtained 
by the presence of fluxes \cite{BlancoPillado:2009di,Carroll:2009dn}. The properties of these vacua are typically greatly varied. For example,  the number of large dimensions  can vary from one vacuum to another \cite{BlancoPillado:2009mi}; a transdimentional landscape. In some cases, one can identify the instanton solutions that represent quantum mechanical
 transitions between these vacua in a process similar to vacuum decay. These
 instantons show that the landscape would be populated by the formation of bubbles 
 of different states. The result 
 is a complex higher dimensional version of an eternal multiverse \cite{BlancoPillado:2009mi}. 
 Our universe can then be within one of these bubbles. 

It is interesting to explore the possibility that we can obtain an observational signature 
of extra dimensions due to their dynamics in our past. However, the prospect of observing these effects will be somewhat hindered by the necessity of 
having a period of slow roll inflation in our immediate past. This is the same situation that 
one faces in any model that tries to explore the pre-inflationary state. Intriguingly, there 
are some anomalies in the CMB data \cite{Adam:2015rua} that might be giving us a 
hint that the duration of inflation was around the minimum required to solve the cosmological problems, but not longer \cite{Cicoli:2014bja}. In such a case, we might hope to see the effects of a previous state 
of the universe in the power spectrum of 
perturbations \cite{Contaldi:2003zv,Bousso:2013uia,White:2014aua,Blanco-Pillado:2015bha,Liu:2013kea,Piao:2003zm}.

There are several different possible transitions. For example, the previous universe could have been effectively lower-dimensional and  the nucleation process 
created a four-dimensional anisotropic bubble. These kinds of transitions and their effects have 
been investigated in several papers \cite{BlancoPillado:2010uw,Graham:2010hh,
Adamek:2010sg,Scargill:2015bva,Blanco-Pillado:2015dfa}, with their main focus on the
anisotropic nature of the initial state of inflation.

Another possibility is  that our universe was the result of  dynamical 
compactification from a higher dimensional universe. The form of the instantons that interpolate 
between highly symmetric vacua were discussed in  Ref.~\cite{BlancoPillado:2009mi}. Investigating the 
spectrum of perturbations of such transitions is quite difficult due to the complicated geometry 
of the transition.

In this paper we simplify the process of the transition, and assume a global 
dynamical compactification of the spacetime.
We assume the spacetime is divided into two parts, the (3+1)-dimensional FRW large dimensions and an internal manifold  of $(d-3)$ flat compact dimensions. Within this model, we consider the existence of an initial period of anisotropic cosmology where the dynamics are controlled by a higher dimensional
fluid source. We then assume that  a compactification mechanism causes the universe to quickly become effectively four-dimensional  and undergoes a period of  inflation of around 60 e-folds. This limited amount of inflation allows us to investigate the  
state of the scalar field that controls the cosmological perturbations; which is an excited state compared to the standard Bunch Davies vacuum.
We compute the spectrum of perturbations from this excited state, compute the multipole moments in the CMB, and compare to data.

This is a simplified toy model since we do not provide a detailed compactification
process. Our focus in this paper is to explore the effects that a rapid compactification
would have on the CMB, and we hope this simple model captures some of the key effects 
that one can expect from more realistic situations. Similar approaches to the one presented here, albeit in a different context, include \cite{Garriga:1989jx,Gasperini:1992sm,Demianski:1993nq,Biesiada:1994pu}.

This  paper is organized as follows.   In Section \ref{sec:Model} we present the toy model for dynamical compactification  and describe the constraints on the energy momentum
tensor necessary to lead to this cosmological history. In  Section \ref{sec:Perturbations} we compute the  spectrum of cosmological perturbations.  In Section \ref{sec:CMB} we compute the observational predictions for the CMB using the CLASS package. Finally,  in  Section \ref{sec:FIN} we conclude.

\section{The model}\label{sec:Model}

\subsection{The metric}

We model a transition from a higher dimensional cosmological stage where the internal dimensions were dynamical, to a purely 4d inflationary period where the 
degrees of freedom associated with the extra-dimensions are fixed. In order to do that we will
postulate a universe (with $d$ spatial dimensions) that is described by a $(d+1)$-dimensional anisotropic metric with two 
scale factors $a(\eta)$ and $b(\eta)$ where $\eta$ is the conformal time i.e. $a(t) d\eta =dt$,
\bel{metricFull}
	ds^2= a(\eta)^2 \(-d \eta^2 + \sum_{i=1}^3dx_i^2 \) + b(\eta)^2 \sum_{j=1}^{d-3}dy_j^2 \,,
\ee
where, for simplicity we have taken the internal space to be a flat (d-3)-dimensional torus. We will also consider that the
higher dimensional theory is controlled by Einstein's equations in a $(d+1)$-dimensional spacetime.
Finally, we also need to specify the matter content of the theory that will determine the dynamics of the
spacetime. In the past this has been a very active area of research and several models with different 
sources have been presented. One can imagine that some of the fields on the higher dimensional
theory play the role of the sources \cite{Freund:1982pg} but there could also be other ingredients present such as
higher dimensional perfect fluids \cite{Sahdev:1985ye}, extended objects \cite{Brandenberger:1988aj}, or 
even sources due to quantum effects \cite{Candelas:1983ae}.

Here we will take a more model independent approach and analyze a family of simple 
metrics of the form described above. The type of metrics
that we will consider for the two stages before and after compactification will be of the following form.

\subsubsection{Higher dimensional evolution.}

Before the transition we have the scale factors given by,
\bel{scalefactoransatz}
a(\eta)= {1\over {(-H \eta_0)}}\left({{\eta}\over{\eta_0}}\right)^{\alpha}~, \qquad b(\eta)=  b_0 \left({{\eta}\over{\eta_0}}\right) ^{\beta} \,,
\ee
which is valid for the range $-\infty < \eta < \eta_0$, where we take $\eta_0 < 0$ to be the time at which
the compactification mechanism dominates and the universe becomes four-dimensional. 
Furthermore $\alpha$ and $\beta$ control the expansion or contraction of the $(3+1)$ as
well as the internal manifold. The rest of the parameters are fixed by the subsequent $4d$ evolution and
the requirement that our scale factors are continuous across the transition.

\subsubsection{Four-dimensional inflation, after the transition}

After the compactification we assume the universe enters a pure $4d$  expansion controlled
by an effective de Sitter space, so the scale factors in this case become,
\be
a(\eta)= {1\over {(-H \eta)}}~, \qquad b(\eta)=  b_0  \,.
\ee
where $H$ is the Hubble constant associated with the $4d$ de Sitter spacetime and $b_0$
denotes the factor that fixes the size of the internal space.

\subsection{Particular examples of the higher dimensional pre-inflationary stage}

The family of solutions we described above encapsulates some interesting cases that 
are worth mentioning explicitly for future reference.

\subsubsection{Higher dimensional de Sitter space}

If we take $\alpha = \beta = -1$ and fix $b_0 = (-H \eta_0)^{-1}$ we have a symmetric situation.
The universe expands isotropically in this higher dimensional stage and in fact it represents
a $(d+1)$-dimensional de Sitter space. This could represent one of the multiple vacua that exists
in models of flux compactification recently discussed in the literature \cite{BlancoPillado:2009di,
Carroll:2009dn}. If so, transitions from this spacetime to the 4d inflationary case can be used to estimate the effects to be expected in these
kinds of transitions \cite{BlancoPillado:2009mi,Carroll:2009dn}.

\subsubsection{Kasner Solutions.}

Another simple set of solutions that are captured by our ansatz are the vacuum solutions \cite{Chodos:1979vk} given by
\be
\alpha=\frac{1}{2} \left(1-\sqrt{3-\frac{6}{d-1}}\right) \,,
\ee
and
\be
\beta=\sqrt{\frac{3}{3-4d+d^2}} \,.
\ee
Here we take the branch of solutions that correspond to a $4d$ expanding
universe \footnote{The exception is the $d=4$ case  where the space time is a higher 
dimensional generalization of the $2d$ Misner spacetime where the four-dimensional part is 
static while the internal circle collapses. There is also the family where the large dimensions 
are contracting but we will not consider them here.}. The internal space is collapsing and one 
can show that all these solutions tend towards a singularity. These are nothing more than the 
higher dimensional generalizations of the familiar Kasner solutions in $4d$.  On the other hand,
the expectation here is that the higher dimensional energy momentum tensor needed for
compactification would change the behavior of the solution before such singularity arises
and the transition to an inflationary $4d$ universe will take place.

\subsection{Constraints on the background evolution}

As we mentioned earlier, we expect many different ingredients to possibly contribute
to the effective energy-momentum tensor that compactifies the spacetime. This is the main 
reason to parametrize our lack of knowledge by a generic ansatz. Nevertheless we want 
to restrict ourselves to physical models where the
total energy is positive and where the equation of state for the effective fluid is such that $-1 \leq w \leq1$.
These constraints limit  significantly the range of the parameters in our set of models.

Using Einstein's equations and the metric in \eqref{metricFull} we can compute the properties of the required energy momentum tensor, by first computing the Einstein tensor,
\bea
G^0_0&=&\frac{(d-3) a b \left(2 \left(a^4+2\right)  a'  b'-a \left(a^4-1\right)  b''\right)-6 b^2  a'^2-(d-4) (d-3) a^6  b'^2}{2 a^4 b^2}~,\nn
G_i^i&=&-\frac{(d-3) a^2 b \left(\left(a^4+1\right)  b''-2 a^3 a'  b'\right)+b^2 \left(6 a'^2-4 a a''\right)+(d-4) (d-3) a^6  b'^2}{2 a^4 b^2}~, \nn
G_j^j&=&-\frac{a b \left((d-5) a^4+d-3\right) \left(a  b''-2  a' b'\right)+6 b^2 \left(2 a'^2-a a''\right)+(d-5) (d-4) a^6  b'^2}{2 a^4 b^2}~,
\eea
where the prime denotes differentiation with respect to conformal time. For the special case $a(\eta) =- (\eta/\eta_0)^\alpha/H \eta_0$ and $b(\eta)=b_0 (\eta/ \eta_0 )^\beta$ becomes,
\beal{EOS}
G^0_0&=&-\frac{\eta_0^2 H^2}{2\eta^2} \left(\frac{\eta }{\eta_0}\right)^{-2 \alpha } \left(6 \alpha ^2+6 \alpha  \beta  (d-3)+\beta ^2 (d-4) (d-3)\right)~,\nn
G_i^i&=&-\frac{\eta_0^2 H^2}{2\eta^2} \left(\frac{\eta }{\eta_0}\right)^{-2 \alpha } \left(2 (\alpha -2) \alpha +2 (\alpha -1) \beta  (d-3)+\beta ^2 (d-3) (d-2)\right)~, \nn
G_j^j&=&-\frac{\eta_0^2 H^2}{2\eta^2} \left(\frac{\eta }{\eta_0}\right)^{-2 \alpha } \left(6 \alpha ^2-6 \alpha +4 \alpha  \beta  (d-4)+\beta  (d-4) (\beta  (d-3)-2)\right)~.
\eea
The equations of state for the four-dimensional and extra dimensional fluids are then given by
\bea
w_{\rm 4D}&=&-\frac{2 (\alpha -2) \alpha +2 (\alpha -1) \beta  (d-3)+\beta ^2 (d-3) (d-2)}{6 \alpha ^2+6 \alpha  \beta  (d-3)+\beta ^2 (d-4) (d-3)}~, \nn
w_{\rm ext} &=&-\frac{6 \alpha ^2-6 \alpha +4 \alpha  \beta  (d-4)+\beta  (d-4) (\beta  (d-3)-2)}{6 \alpha ^2+6 \alpha  \beta  (d-3)+\beta ^2 (d-4) (d-3)}~,
\eea
where $w_{\rm 4D}$ is the $4d$ part and $w_{\rm ext}$ corresponds to the analogous quantity
taking into consideration the extra-dimensional part of the pressure. 
 
\begin{figure}
\center
\includegraphics[scale=0.7]{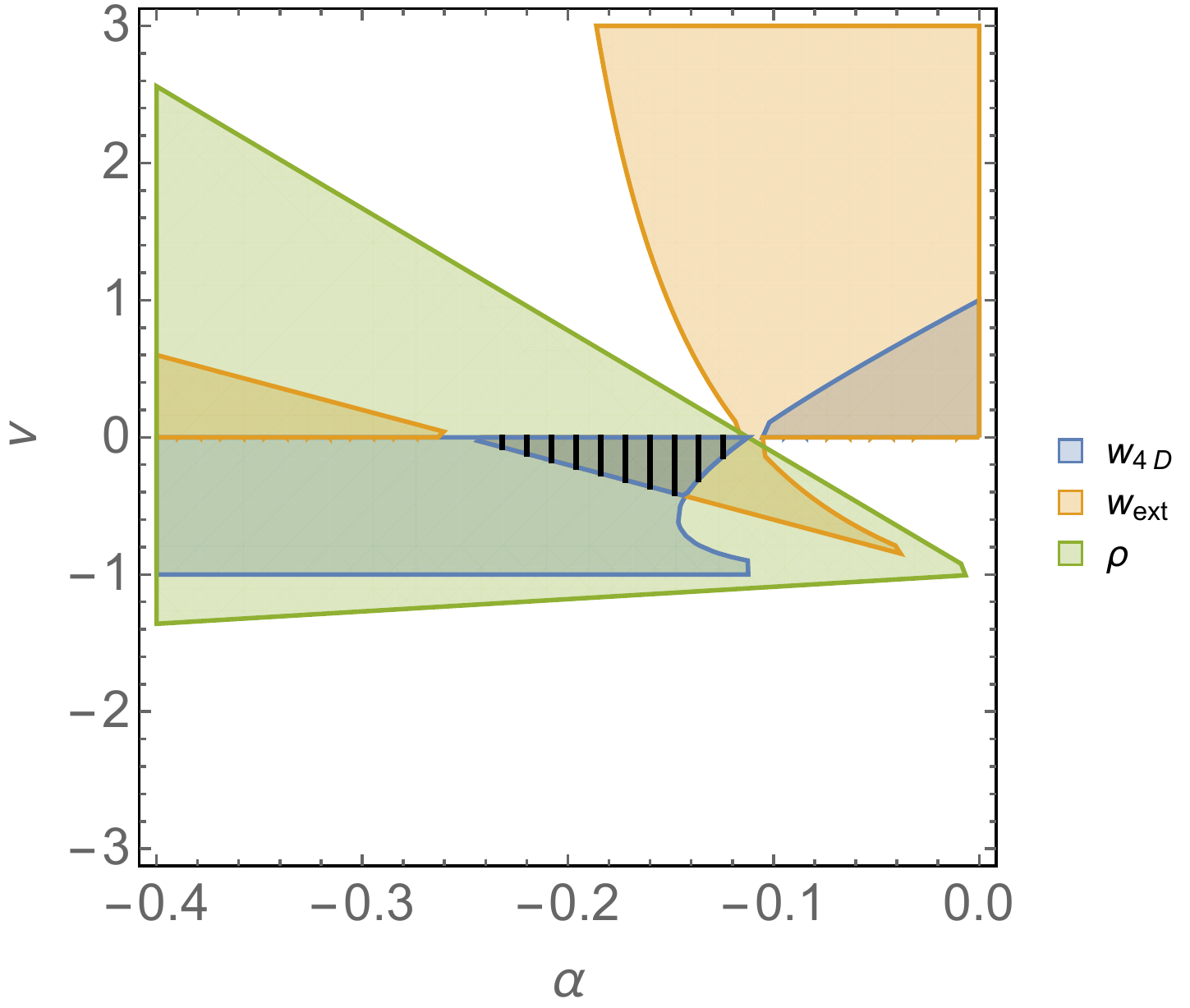}
\caption{The shaded regions shows the allowed values for $\alpha$ and $\nu$ for the different constraints. The overlapping hatched region in the center of the figure represents the parameter space that respects all the constraints. \label{fig:alphanu}}
\end{figure}

 Considering all these requirements the allowed region of the $(d, \alpha, \beta)$ parameter space 
gets reduced significantly. We show in Fig. (\ref{fig:alphanu}) the space allowed by these constraints as the
overlap of all the shaded regions for the case $d=4$, analogous situations can be found for 
other dimensions.
Here we have expressed $\beta$ in terms of
\be\label{nuDef}
\nu=-1+2\alpha+(d-3)\beta \,,
\ee
for later convenience. We also represent only the region with $\alpha<0$ since these would be
the cases that interest us in this paper.

\section{Scalar Field Perturbations}\label{sec:Perturbations}

Here we will consider the cosmological perturbations generated in the spacetime described before. In order to do
that we will first consider the perturbations for a massless scalar field in this background with action,
\beal{eq:Lagrangian}
S&=&-\frac12\int d \eta dx^3 dy^{d-3} \sqrt{-g}  \partial_\mu \phi \partial^\mu \phi~.
\eea
The equation of motion for this field in a metric given by Eq. \eqref{metricFull} is
\be
\Box \phi= \phi'' + \(\frac{2 a'}a+ \frac{(d-3) b'}{b} \)  \phi' - \sum_{i=1}^d\partial_i \partial_i \phi- \frac{a^2}{b^2}\sum_{j=d+1}^{d-3}\partial_j \partial_j \phi=0.
\ee
Expanding the field in terms of Fourier modes, 
\be
\phi(\eta) = \int{{{d^3k}\over{(2 \pi)^{3/2}}} {{d^{d-3}k_y}\over{(2 \pi)^{(d-3)/2}}}\left[ a_{(k, k_y) } \varphi_{(k, k_y) }(\eta)  e^{i  k \cdot x} e^{ik_y \cdot y} + \text{cc} \right]}~,
\ee
we arrive to the equations for the mode functions,
\be
\varphi'' + \(\frac{2 a'}a+ \frac{(d-3)  b'}{b} \)  \varphi' + \left(k^2+\frac{a^2}{b^2} k_y^2 \right) \varphi=0 \,.
\ee
The field in \eqref{eq:Lagrangian} does not 
have a canonical kinetic term. We can define a canonically normalized field  $v(\eta)$  by the field redefinition,
\be
\varphi (\eta)= a(\eta)^{-1} b(\eta)^{-(d-3)/2}  ~v(\eta)~,
\ee

which leads to 
\bel{newEOM}
	 v'' - \( \frac{ a''}{a}+\frac{(d-3) a'  b' }{ab}+\frac{(d-3)  b''}{2 b}+\frac{(d-3)(d-5)   (b')^2}{4 b^2}-\frac{a^2}{b^2}k_y^2-k^2 \)v=0 \,.
\ee

Taking the general ansatz for the scale factors in 
Eq. \eqref{scalefactoransatz}, one arrives at the equation
\bea \label{abEOM}
v'' - \bigg[ \frac{\alpha(\alpha-1)}{\eta^2}+\frac{d-3}{2\eta^2}\[2 \alpha\beta +\beta(\beta-1)+\frac{(d-5)\beta^2}{2}\]  - {1\over {(-H \eta_0 b_0)^2}}\left({{\eta}\over {\eta_0}}\right)^{2(\alpha-\beta)} k_y^2-k^2 \bigg]v=0~. \nonumber
 \eea
 Let us first consider the evolution of the purely four-dimensional modes 
 \footnote{We will later comment on the relevance of the Kaluza-Klein massive modes.}, those with $k_y=0$. In this case, 
one can get a general solution of this equation of the form,
\be
v_k(\eta)=\sqrt{|k\eta|} \left(A_k J_{\nu/2}(|k\eta|)+B_k Y_{\nu/2}(|k\eta|)\right) \,,
\ee
where $\nu$ is defined in Eq.~\eqref{nuDef}.
Fixing the coefficients $A_k$ and $B_k$ one identifies the vacuum state for these mode functions. The usual 
approach in these situations, in particular in an inflationary stage, is to identify the mode functions with the ones given
by the so-called Bunch-Davies vacuum. Hence, at very early times, where the modes are deep inside the horizon, they should match to the Minkowski modes, therefore
\be
\lim_{\eta \rightarrow -\infty} v_k(\eta)= \frac{1}{\sqrt{2k}}e^{-ik\eta} \,.
\ee

This is a legitimate assumption if the $4d$ comoving Hubble radius decreases with time so that four-dimensional perturbations start being sub horizon and then leave the horizon as time passes.  In our ansatz 
this is the case when $\alpha < 0$. In the following  we will only consider this possibility.

Using the asymptotic form of the Bessel functions of type $J$ and $Y$ for $k|\eta| \gg 1$  
\bea
	J_{\nu/2}(z) \approx \sqrt{2\over \pi |z|} \cos\(|z|-\frac{\nu \pi}{4} - \frac\pi4\)~, \nn
	Y_{\nu/2}(z) \approx \sqrt{2\over \pi |z|} \sin\(|z|-\frac{\nu \pi}{4} - \frac\pi4\)~,
\eea
we can compute the coefficients $A_k$ and $B_k$ that match correctly at early times.
\be 
	A_k= -i B_k= \sqrt{\pi \over 4 k}e^{i \pi(1+\nu)/4} \,.
\ee
Therefore the modes are given by 
\bel{eq:modes}
v_k(\eta) =\frac{\sqrt{\pi |\eta|}}2 e^{i \pi(1+\nu)/4} \[ J_{\nu/2}(k |\eta|)+ i  Y_{\nu/2}(k|\eta|)\] =\frac{\sqrt{\pi |\eta|}}2 e^{i \pi(1+\nu)/4}H_{\nu/2}^{(1)}(k |\eta|) ~.
\ee
Here $H_{\nu/2}^{(1)}$ are Hankel functions of the first kind. All the information about the background evolution is 
therefore encoded in the index of the  Hankel function, $\nu$. In particular this also gives us the information 
about the spectral index of the power spectrum of the perturbations that leave the horizon during this period. Using their limiting form 
\be
H^{(1)}_{\nu/2}(x)= \frac{i 2^{d/2} x^{-\frac{\nu}{2}} \Gamma \left(\frac{\nu}{2}\right)}{\pi }~,
\ee
we get the power-spectrum: 
\be
	P(k)= \lim_{k \eta\rightarrow 0} a^{1-\nu} |v_k|^2=\lim_{k |\eta| \rightarrow 0} (H \eta)^{\nu-1} |v_k|^2= H^{\nu-1} 2^{\nu-2} \pi^{-1} k^{-\nu} \Gamma\(\frac{\nu}{2}\)^2~.
\ee

For future reference we note that for our particular examples we have,

\be
\nu_{dS_D} = -  d ~~~~~~~~~\text{for a $(d+1)$ de Sitter spacetime, so the spectrum goes to $k^{-d}$}
\ee
and 
\be
\nu_{\text{Kasner}} = 0 ~~~~~~~~~~~\text{for the vacuum solutions for any $d$ so the spectrum is $k$ independent.}
\ee

\subsection{The Compactification Transition}

As we mentioned in the previous section, we would like to understand the effect on the spectrum of
perturbations of a transition from a higher dimensional universe to a purely $4d$ de Sitter space.
We model this transition as a quick change in scale factors at a particular
time $\eta=\eta_0$.

After the transition the Eq.~(\ref{abEOM}) becomes
\be
v'' + \left(k^2 -\left(2 - \frac{m_y^2}{ H^2}\right) \frac{1}{\eta^2} \right)v =0~,
\ee
where we have denoted by $m_y$ the masses of the Kaluza-Klein states given by 
\be
m_y = {{k_y}\over {b_0}}~.
\ee
The general solution is $w$ instead of $v$
\be
w_{k,k_y}=\sqrt{\eta} \left[C_{k,k_y} J_{\mu/2}(k|\eta|)+D_{k,k_y} Y_{\mu/2}(k|\eta|)\right] \,,
\ee
where
\be
\mu=\sqrt{9-4\left(\frac{m_y}{ H}\right)^2} \,.
\ee
Considering the mode functions for the zero modes, we should take $k_y=0$, so $\mu=3$ and the solutions become
\bel{eq:modesAfter}
	w_k(\eta) =\sqrt{2 \over \pi k} \[ C_k\( \frac{\cos k \eta}{k \eta }+ \sin k \eta \)+ D_k \( \cos k\eta - \frac{\sin k\eta }{k\eta }\)\] ~,
\ee
which is the usual four-dimensional result for a massless scalar during inflation.

We now have the expression for the mode functions before and after the transition so
the only thing left to do is to match these solutions across the surface $\eta=\eta_0$. The mode functions must be continuous across this boundary, namely,
\bel{eq:modeMatch}
	w(\eta_0)= v(\eta_0)~. \nn
\ee
On the
other hand we can not impose the same continuity for the derivative of the mode functions. The reason for 
this can be found by looking at Eq. \eqref{newEOM}. In our model the transition between the two stages occurs by 
a sharp variation of the scale factor. This means that the derivative of $a(\eta)$ and $b(\eta)$ are discontinuous
across the transition and therefore its second derivative would have a delta function contribution. For our particular
case we have,
\be
{{a''}\over{a}} \supset - \left({{1+ \alpha}\over {\eta_0}} \right)\delta(\eta - \eta_0) \,~, 
\ee
and 
\be
{{b''}\over{b}} \supset - \left({{\beta}\over {\eta_0}} \right)\delta(\eta - \eta_0) \,.
\ee

Therefore integrating Eq. \eqref{newEOM} across the surface of matching on obtains that 
the derivative of mode functions $w$ and $v$ acquires a jump given by,
\be \label{deriv}
 w' (\eta_0)-  v' (\eta_0)= -\frac{(3+\nu)}{2 \eta_0} v(\eta_0)~.
\ee
From these matching conditions we can calculate the coefficients $C_k$ and $D_k$ in mode function $w(\eta)$ which are given by
\bea
C_k&=&-\frac{\pi  e^{\frac{1}{4} i \pi  (\nu+1)}}{2 \sqrt{2} \(-\eta _0 k\)^{1/2}}\Big\{\eta _0 k \sin \(\eta _0 k\) H_{\frac{\nu }{2}}^{(1)}\!\(-k \eta _0\)+H_{\frac{\nu +2}{2}}^{(1)}\(-k \eta _0\) \[\eta _0 k \cos \(\eta _0 k\)-\sin \(\eta _0 k\)\]	\Big\}~,\nn
	D_k&=&\frac{\pi  e^{\frac{1}{4} i \pi  (\nu+1)}}{2\sqrt{2} \(-\eta _0 k\)^{1/2}}\Big\{-\eta _0 k \cos \(\eta _0 k\) H_{\frac{\nu}{2}}^{(1)}\!\(-k \eta _0\)+ H_{\frac{\nu +2}{2}}^{(1)}\(-k \eta _0\) \[\eta _0 k \sin \(\eta _0 k\)+\cos \(\eta _0 k\)\]\Big\}~.\nn
\eea

These determine the mode function $w(k)$. From those functions we can get the power spectrum for the scalar field after the transition as, 
\bea \label{powerspectrumA}
P_\phi(k)&=&\lim_{\eta\to 0} (H\eta)^2 |w_k(\eta)|^2 \\
&=& \frac{\pi  H^2}{4 |\eta _0| k^4} \bigg| \eta _0 k H_{\frac{\nu }{2}}^{(1)}\left(-k \eta _0\right) \sin k\eta _0+H_{\frac{\nu +2}{2}}^{(1)}\left(-k \eta _0\right) \left(\eta _0 k \cos k\eta _0-\sin k\eta _0\right) \bigg|^2  \,.
\eea

We can normalize this power spectrum to obtain,
\bea \label{powerspectrum}
{\cal P}_\phi (k) = {{2k^3}\over {H^2}} P_\phi(k)~.
\eea
This normalized power spectrum has the property that goes to one in the usual Bunch-Davies
vacuum and deviates from there otherwise, so it is a good way to measure the deviation
or excitation of our power spectrum.

\vspace{0.2in}

\begin{figure}
\center
\includegraphics[scale=0.5]{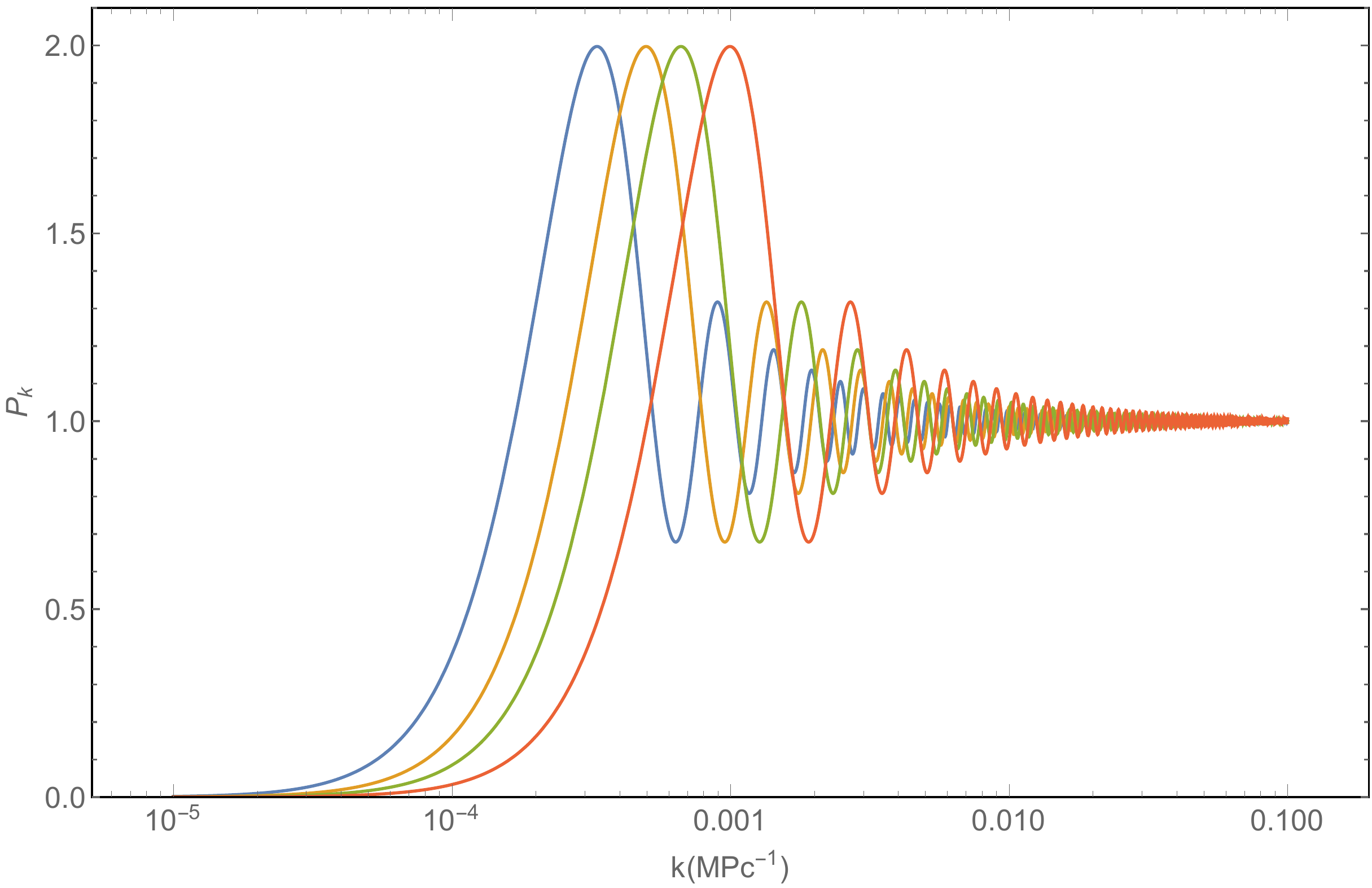}
\caption{ Normalized power spectrum for different values of compactification time $\eta_0$.  For later compactification times the power spectrum is shifted to the right. \label{fig:normal-spectrum-eta}}
\end{figure}

\begin{figure}
\center
\includegraphics[scale=0.5]{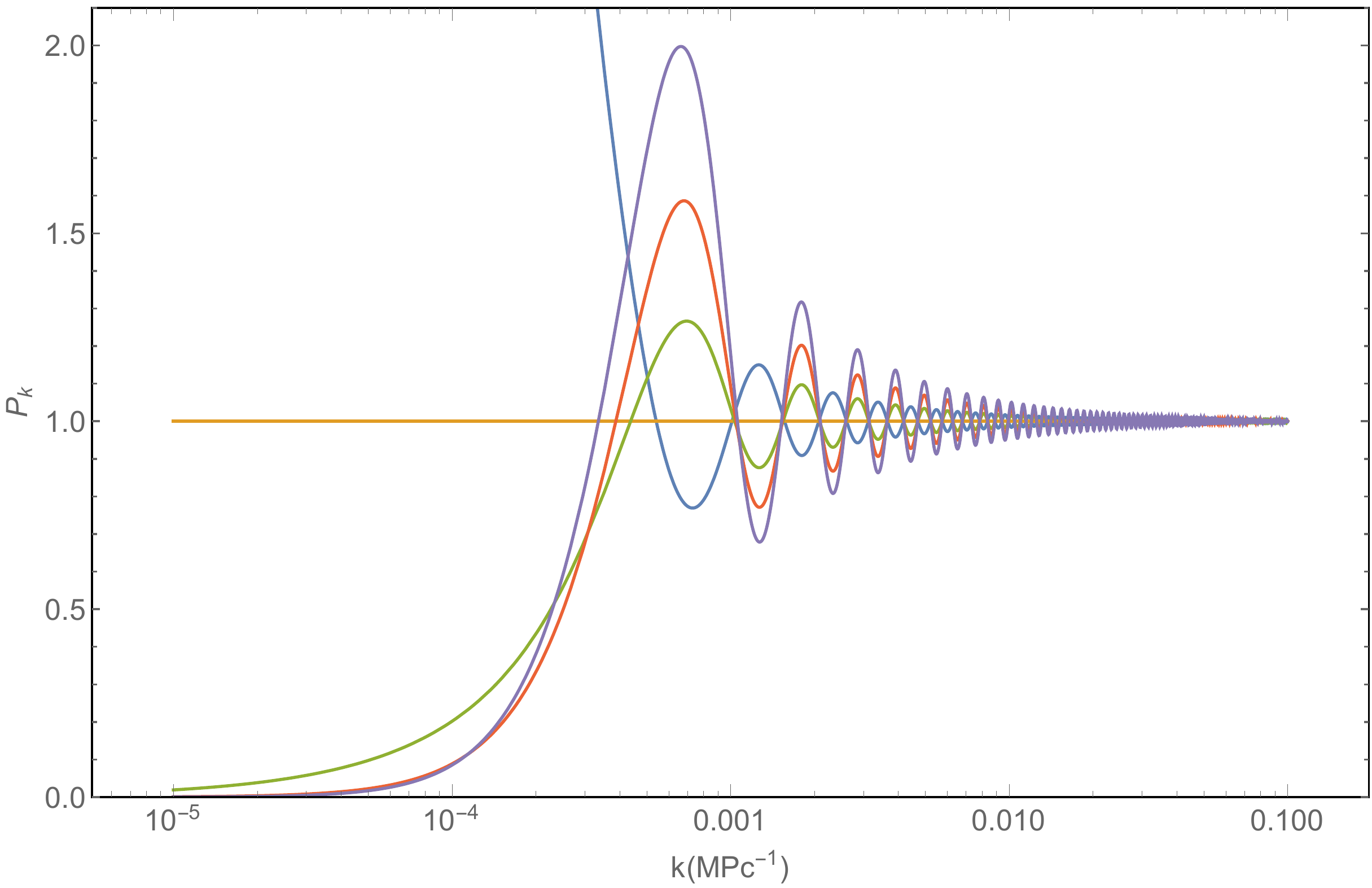}
\caption{Normalized power spectrum for different values of the parameter $\nu$. For $\nu=-4$ (blue) we have enhancement of power for low $\ell$'s. For $\nu=-2$ (green), $\nu=-1$ (orange) and $\nu=0$ (purple) we have suppression of power for low $\ell$'s. For $\nu=-3$ we return to the four-dimensional case.} \label{fig:normal-spectrum-nu}
\end{figure}

This power spectrum is parametrized by two quantities,  $\eta_0$ and $\nu$. We show in Figs.~\ref{fig:normal-spectrum-eta} and \ref{fig:normal-spectrum-nu}
the normalized power spectrum for different values of  $\eta_0$ and $\nu$. The role of 
$\eta_0$ is clear physically; it controls the moment at which the transition takes place.
Changing this number shifts the power spectrum along the $k$-axis so
its value fixes the transition point on the power spectrum graph. Modes that 
leave the horizon well after the transition evolve most of their lives in a pure $4d$ de Sitter
stage and are therefore driven towards a flat spectrum. On the other hand, modes
that left the horizon during the pre-inflationary stage were affected by different
dynamics. The transition from one kind of spectrum to the other occurs at the scale
of the horizon at time $\eta_0$.

The effect of the parameter $\nu$ is two fold. On one hand, it controls the spectral index of the ``initial power spectrum" of the
mode functions; the one they would have if only the first stage of the evolution existed. This 
is easy to see by realizing that $\nu$ appears on the index of the Hankel functions and therefore
on their asymptotic form. On the other hand, $\nu$ appears also in the expression for the
jump of the first derivative of the mode functions. This has as its most dramatic effect
a change on the height of the power spectrum at the transition region, in particular on its
first peak. This is an effect that is clearly specific of our model and in principle it could
set it apart from other similar models.

\vspace{0.2in}

\subsection{Curvature perturbations}
To convert from scalar field modes to physical curvature perturbations, we introduce $\mathcal{R}=-\delta\phi/(\sqrt{2\epsilon}\,M_{Pl})$. Moreover, to convert from modes in pure de Sitter to modes in a slow-roll background, we replace $H\to H_*(k/k_*)^{-\epsilon_*}$ in (35) and replace $\epsilon\to\epsilon_*(k/k_*)^{4\epsilon_*-2\eta_*}$ in the conversion factor to $\mathcal{R}$, as different modes see a different Hubble parameter. This gives  \cite{Riotto:2002yw}
\be\label{curvaturepowerspectrum}
P_{\mathcal{R}}(k) = {\cal P}_{\phi}(k)\times \left( {\cal A}_{\mathcal{R}}\left(\frac{k}{k_*}\right)^{n_s -1 }\right) \,,
\ee
where $n_s-1=-6\epsilon_*+2\eta_*$ and the amplitude is $\mathcal{A}_{\mathcal{R}}=H_*^2/(4M_{Pl}^2\epsilon_*)$ where the subscript $*$ refers to a pivot scale. Here $\mathcal{P}_\phi(k)$ is the normalized  version of the power spectrum computed in  Eq.~(\ref{powerspectrum}), $n_s$ is the spectral index which depends on the slow roll parameters and ${\cal A}_{\mathcal{R}}$ is the amplitude of the power spectrum at pivot scale $k_*$.
We take the values of ${\cal A}_{\mathcal{R}}, k_*~ \text{and} ~n_s$ from the latest results from PLANCK \cite{Adam:2015rua}.

\subsection{The massive modes}

As we mentioned in the previous sections the pre-inflationary phase will also excite the 
massive modes and not only the zero modes. However, any successful compactification
mechanism must be the source of the dominant energy momentum tensor at the time of the transition. This means that 
an important assumption in our model is that the energy stored in these modes at the onset of inflation
must be subleading. On the other hand, these modes are by definition
more massive than the Hubble parameter during the second stage of evolution, during the
$4d$ inflationary part and therefore their energy density will be further diluted by the expansion of the
universe during that time. It is therefore likely that this modes do not play a significant role 
in the kind of models we are studying here.

\section{Effects on the CMB}\label{sec:CMB}

Using the CLASS code (\cite{Blas:2011rf}) and the cosmological parameters recently published by the
PLANCK collaboration \cite{Adam:2015rua} we propagated the power spectrum of Eq.~(\ref{curvaturepowerspectrum}) 
to look at the observational effects of dimensional compactifications on the CMB temperature data.

As we described earlier the power spectrum is parametrized by setting the time of compactification $\eta_0$ and
$\nu$ that describes the type of pre-inflationary dynamics. It is clear that using a very early
compactification time would eliminate any possibility of observing any effect of this transition since it
will push the scale of the transition outside of the CMB window of scales.  On the other hand, it is also clear that this transition can not occur too late
since this would ruin the agreement of the power spectrum with the CMB data currently well fitted by a simple
power law. These considerations force us to limit the range of the time of the transition to be around the 
scale of $k = 0.001 ~{\text Mpc}^{-1}$.

\begin{figure}
\begin{center}
\includegraphics[scale=0.5]{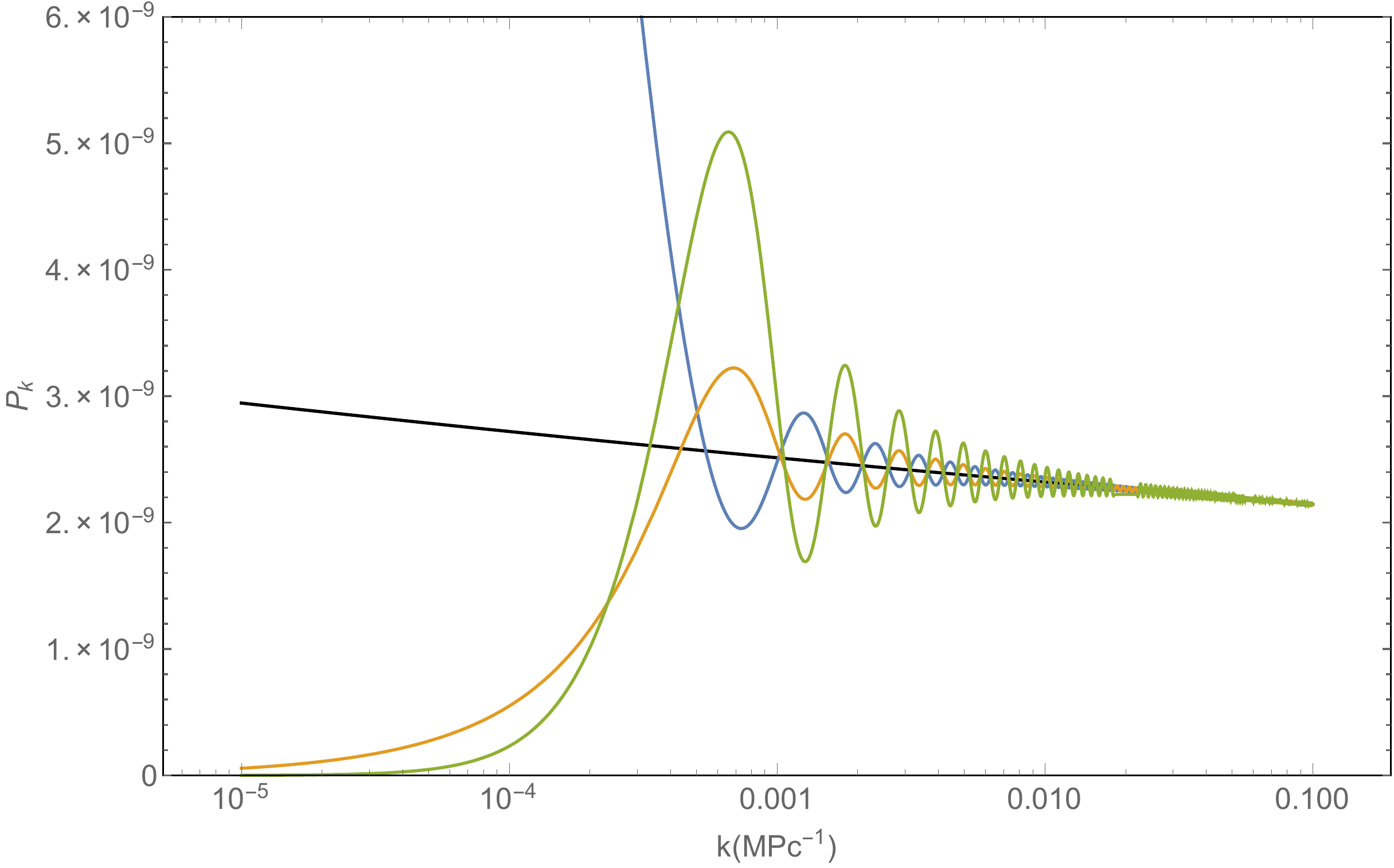}
\includegraphics[scale=0.5]{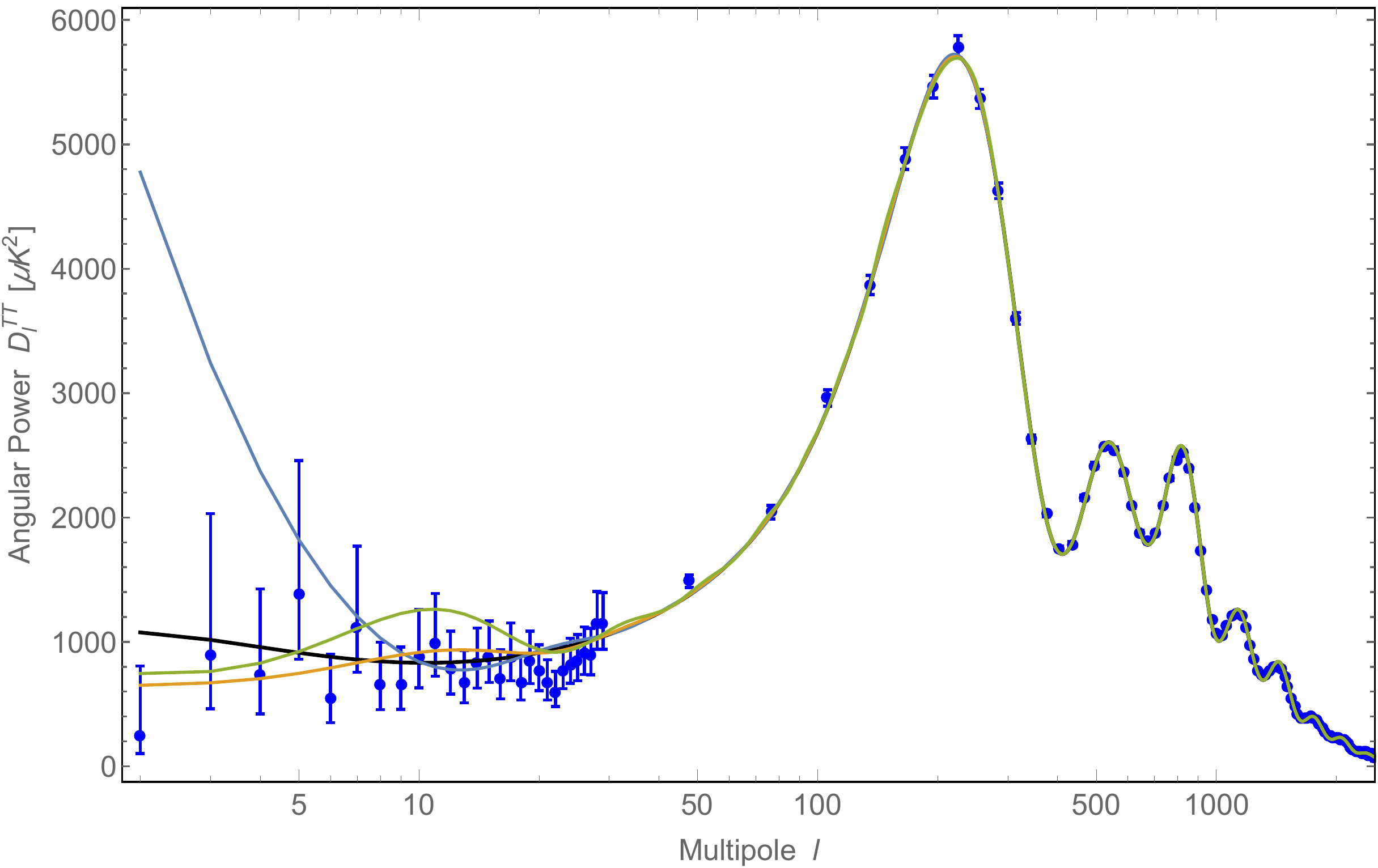}
\caption{Comparison of the PLANCK data points (blue) with the standard  $\Lambda$CDM model (black) and our model for: $\nu=-4$ (blue), $\nu=-2$ (orange) and $\nu=0$ (green).\label{fig:nu}} 
\end{center}
\end{figure}

\begin{figure}
\begin{center}
\includegraphics[scale=0.5]{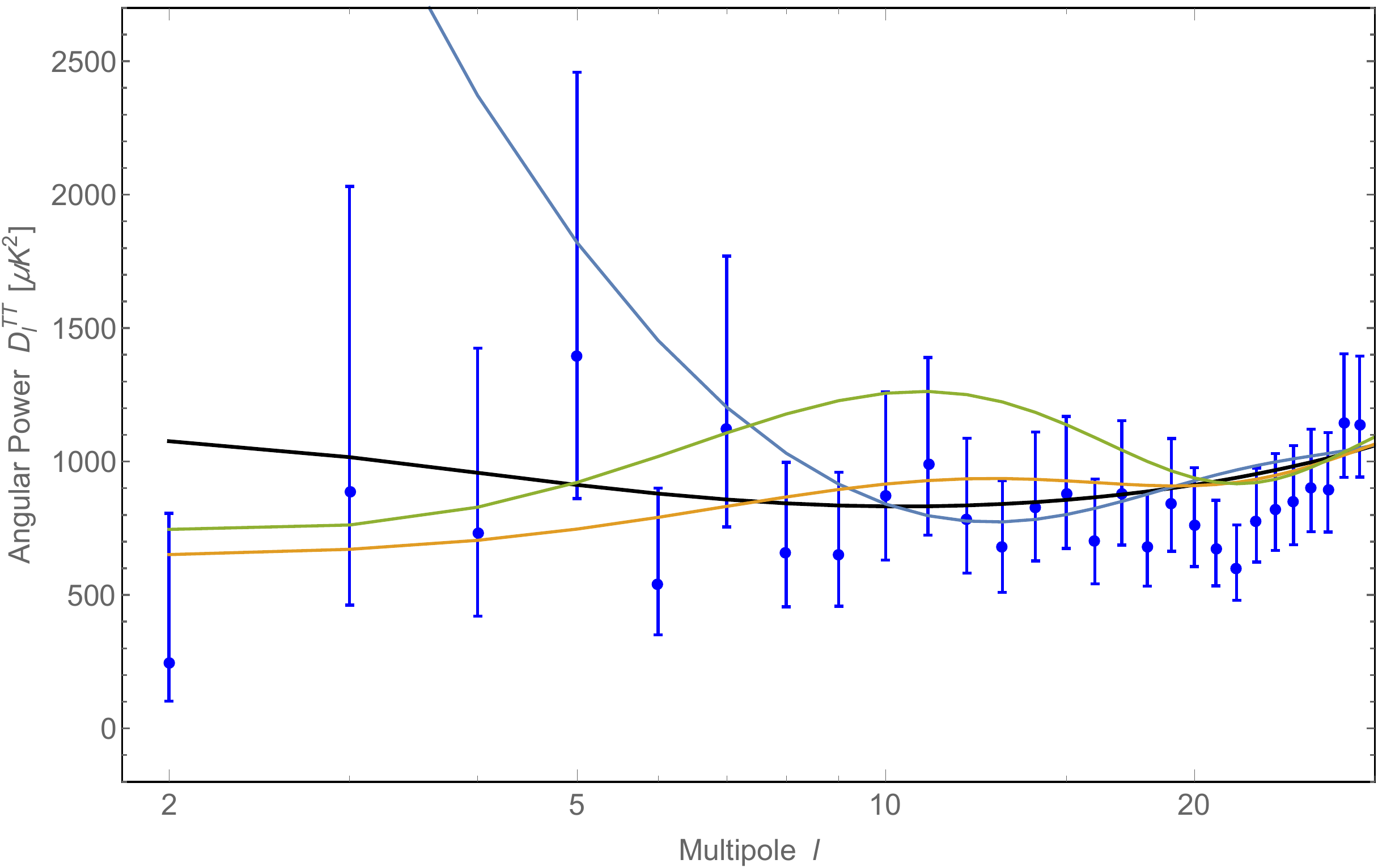}
\caption{Low-$l$ region of the temperature power spectrum for the same parameters as in Fig. \ref{fig:nu}. \label{fig:closeup-lowl} }
\end{center}
\end{figure}

Next we investigate the effect of varying the value of $\nu$ keeping fixed the transition time.
We show in Fig.~(\ref{fig:nu}) the corresponding power spectrum for the temperature fluctuations
computed using different values of $\nu$
and compare the results to the PLANCK data. We show in Fig. (\ref{fig:closeup-lowl})
a close up of the low-{\it l} region for the same set of values.

We notice that the power is significantly enhanced for values of $|\nu| > 3$. This includes all the values
that correspond to a higher dimensional de Sitter space prior to the slow roll period. This seems to 
indicate that if such transdimensional transition occurred it was too far in the past and its observational evidence 
is out of our reach.

For $|\nu| <3$ there is a suppression on the power at small $\ell$ as it seems to be
required by the PLANCK data. However, the details of the transition create a high bump
on the power spectrum that makes the fit more problematic. This is particularly clear
in the case of $\nu=0$ where the suppression is present but the effects of the sharp
transition undo this effect in the region of interest.

We have not attempted to make a 
systematic treatment of the data to see what would be the best fit value but things appear to get better
for $-3 < \nu < -2$. On the other hand, these values seem to be at odds with 
the constraints based on the equation of state of the fluid required to create such 
dynamics. See the discussion in Section \ref{sec:Model}.

\section{Conclusions}\label{sec:FIN}
Observing evidence of the existence of extra dimensions would potentially be revolutionary. In this paper we showed if extra dimensions exist and they undergo a dynamical compactification process in the early universe, this could leave observable imprints on the CMB. This is possible if the subsequent $4d$ inflationary period did not last too long. Using a simple model for the higher dimensional evolution we investigated how the different parameters of this scenario would affect the spectrum of perturbations. 

We found that this mechanism could lead to an enhancement or suppression of the fluctuations in the CMB on large scales. Comparison with the Planck data put important constraints on this kind of transition occurring right before the beginning of inflation. In particular we show that a rapid transition from a higher dimensional de Sitter space to our $4d$ inflation period is ruled out unless there is a large number of e-folds after the transition. 

One point of concern is that the data is better fit in a region of parameter space that requires a higher dimensional fluid with an equation of state outside of the range  $-1 \leq w \leq 1$. Moreover, there are a number of other issues that remain to be investigated. Most importantly, this work does not provide a detailed mechanism of compactification but assumes a rapid transition. While we think this captures the basic effects of a dimensional compactification 
transition, a more realistic model is needed to verify the details of our results. In particular, one could argue that a detailed study of realistic compactifications will imply a slow transition. This will change the details of the power spectrum and likely making some of the models a better fit to the data. 

Finally, it is important to investigate the necessary matter ingredients that are required to produce the transition itself. In particular, one should understand whether a transition of the kind discussed here would imply a violation of any energy conditions. In some cases this is obviously true.  For example, a transition from a collapsing higher dimensional universe to an expanding effectively $4d$ case would violate the null energy condition at the bounce. It would be interesting to see if this is in fact a generic situation for the cases of interest.

\section*{Acknowledgments}

The authors would like to thank Matthew Jonhson for collaboration in the earlier stages 
of this work as well as Larry Ford, Alan Guth,  Alexander Vilenkin and Kepa Sousa for helpful discussions. 
J.J. B.-P. is supported in part by the Basque Foundation for Science (IKERBASQUE), the 
Spanish Ministry MINECO  grant (FPA2015-64041-C2-1P) and  Basque Government grant (IT-979-16). A.M. is supported by a National Science Foundation grant 1518742.

\bibliography{biblio}

\end{document}